\documentclass[14pt, a4paper]{article}

\usepackage{arxiv}

\usepackage{amsmath}
\usepackage{amssymb}
\usepackage{amsfonts}
\usepackage{graphicx}
\usepackage{tikz}
\usepackage{circuitikz}
\usepackage{pgfplots}
\usepackage{geometry}
\usepackage{booktabs}
\usepackage{setspace}
\usepackage{caption}
\usepackage{subcaption}
\usepackage{cite}
\pgfplotsset{compat=1.18}

\title{Electromagnetic Characterization of Magnetic Bar: Case of Square Cross-Section Shape}

\author{%
  Taha El Hajji*,\quad Bruno Ricardo Marques,\quad Lars Sjöberg \\
  Alvier Mechatronics AB, Helsingborg, Sweden \\
  \texttt{taha.elhajji@gmail.com}
}

\begin{document}

\maketitle

\begin{spacing}{1.8}

\begin{abstract}
This paper presents a complete two-dimensional theoretical model for the electromagnetic behavior of square-section solid magnetic bars under sinusoidal loading. Through the application of Maxwell's equations within a Cartesian coordinate system and the integration of complex permeability, exact mathematical expressions are derived for mutual impedance, internal magnetic fields, flux, and core losses. Hyperbolic functions are utilized to separate the variables, enabling the accurate representation of edge flux accumulation and the 2D skin effect. In addition to mathematically decoupling eddy current and hysteresis losses, this formulation yields a new apparent permeability parameter. This parameter establishes a fast, reliable method for magnetic steel characterization that bypasses the extensive processing times associated with Finite Element Analysis (FEA). Numerical results over 1 Hz–1 MHz show the apparent relative permeability decreasing from 500 to 300 and a characteristic resistance peak near 700 kHz, marking the transition from volumetric to surface-dominated loss regimes.
\end{abstract}
\end{spacing}


\clearpage

\section{Introduction}
The performance of modern electromagnetic machinery depends heavily on the precise evaluation of soft magnetic components. Traditional material testing for magnetic steels relies on straight single-sheet or bar apparatuses outfitted with excitation and sensing windings. While one-dimensional mathematical models leveraging Bessel functions are well-established for cylindrical geometries \cite{BarC}, as well as for ring cores featuring circular cross-sections \cite{ArxivTahaRC,ieee668058, ieee5382736}, such 1D simplifications break down for other standard cross-sectional geometries. In particular, custom testing frameworks and laminated cores frequently utilize a square cross-section, which demands advanced multi-dimensional modeling.
As operating frequencies increase and the skin effect emerges, induced eddy currents repel the magnetic flux outward from the core. Within a straight bar possessing a square cross-section, this outward shift causes severe two-dimensional flux accumulation at the geometric extremities, resulting in a drastically non-uniform internal field. Furthermore, incorporating a complex permeability formulation is essential to properly capture the hysteresis-driven phase delay between the applied field and this localized flux.
Driven by these challenges, this paper modifies an established 2D Cartesian analytical technique—previously utilized for toroidal rings with square cross-sections \cite{ArxivTahaRS,ieee4527040,RingS3,RingS4,RingS5}—for application to straight bar topologies. By applying the separation of variables to solve Maxwell’s equations, this study derives precise, closed-form expressions for decoupled core losses, mutual impedance, flux distributions, and internal magnetic fields. Ultimately, this mathematical model delivers a rapid, highly rigorous alternative for computationally demanding Finite Element Analysis (FEA).

\section{Nomenclature}
\begin{itemize}
    \item $L$: Total length of the solid magnetic bar [m]
    \item $l_1$: Length of the primary excitation winding [m]
    \item $a$: Side length of the square cross-section [m]
    \item $A$: Cross-sectional area, calculated as $A = a^2$ [m$^2$]
    \item $\mu_c$: Complex magnetic permeability [H/m]. Expressed as $\mu_c = \mu' - j\mu''$
    \item $\mu'$: Real part of complex permeability representing energy storage [H/m]
    \item $\mu''$: Imaginary part of complex permeability representing energy loss [H/m]
    \item $\sigma$: Electrical conductivity of the core material [S/m]
    \item $\omega$: Angular frequency of the sinusoidal excitation, $\omega = 2\pi f$ [rad/s]
    \item $N_1$: Number of turns of the primary winding
    \item $N_2$: Number of turns of the secondary winding
    \item $I$: RMS current supplied to the primary winding [A]
    \item $V_2$: Voltage induced across the secondary winding [V]
    \item $\mathbf{H}$: Magnetic field intensity [A/m]
    \item $\mathbf{B}$: Magnetic flux density [T]
    \item $\mathbf{E}$: Electric field intensity [V/m]
    \item $Z_m$: Mutual complex impedance of the bar measurement setup [$\Omega$]
    \item $Z_{core}$: Equivalent primary complex impedance of the core [$\Omega$]
    \item $Y_{core}$: Complex admittance of the core, defined as $Y_{core} = G_{core} - jB_{core}$ [S]
    \item $G_e, G_h$: Eddy current and hysteresis equivalent conductances [S]
    \item $P_e, P_h$: Eddy current and hysteresis power losses [W]
    \item $x, y, z$: Local Cartesian coordinates within the square cross-section [m]
    \item $\gamma$: Complex propagation constant, defined as $\gamma = \sqrt{j\omega \mu_c \sigma}$ [1/m]
    \item $k$: Complex separation constant in the core, evaluated as $k = \sqrt{j\omega \mu_c \sigma / 2}$ [1/m]
    \item $\mu_{app}$: Apparent relative permeability accounting for eddy currents [H/m]
\end{itemize}

\section{Modeled Bar and Measurement Setup}

\subsection{Geometry of the Square Bar}
\begin{figure}[h]
    \centering
    \includegraphics[width=0.5\linewidth]{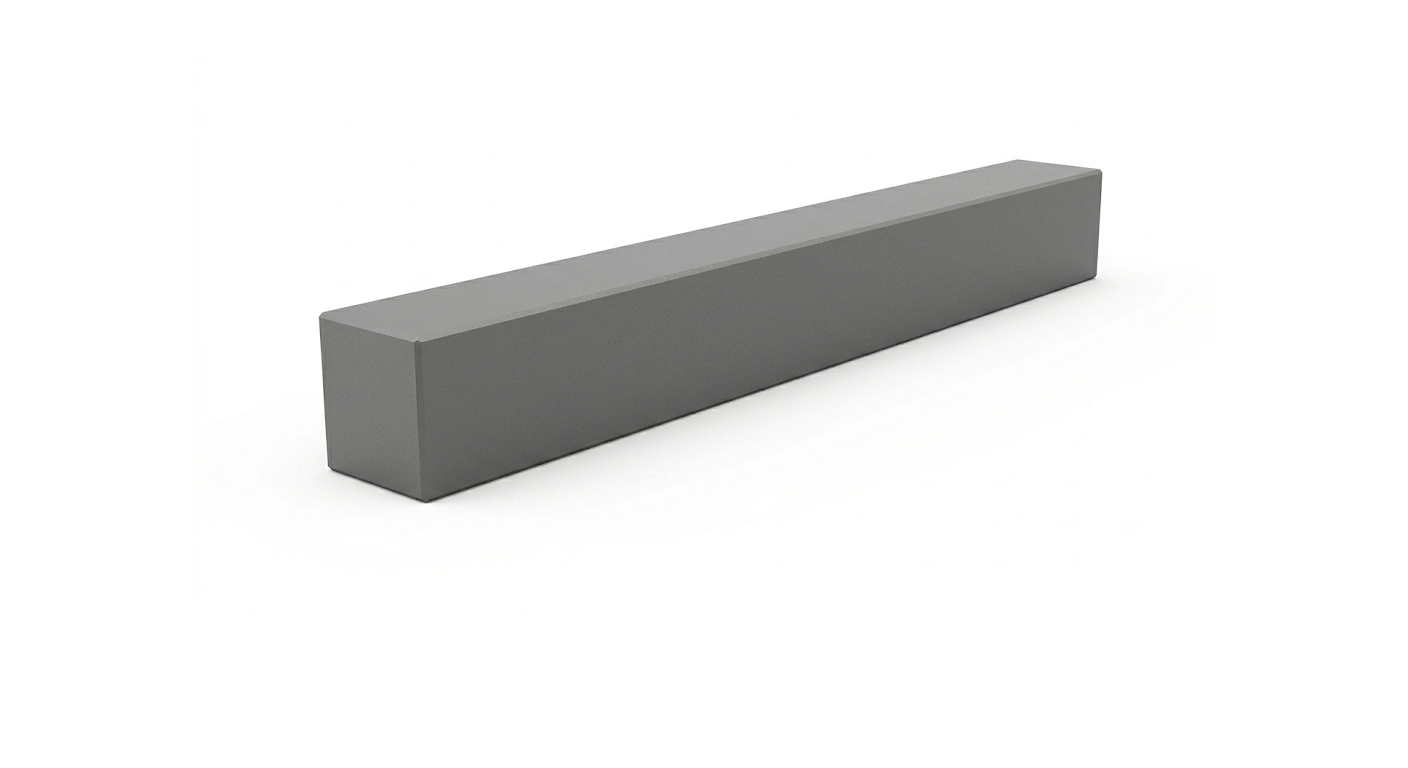}
    \caption{3D view of the studied bar.}
    \label{fig:3Dbar}
\end{figure}

The geometry under consideration is a long, straight magnetic steel bar of length $L$, which has a square cross-section with side dimensions equal to $a$ (see Fig.~\ref{fig:3Dbar}). We introduce a localized Cartesian reference frame $(x,y,z)$ positioned such that its origin $(0,0,0)$ coincides with the center of the 2D profile. In this arrangement, the $z$-axis runs continuously down the length of the bar, thereby restricting the physical domain of the square cross-section to the planar coordinates $x \in [-a/2, a/2]$ and $y \in [-a/2, a/2]$.

\subsection{Winding Configuration}
\begin{figure}[h]
    \centering
    \includegraphics[width=0.5\linewidth]{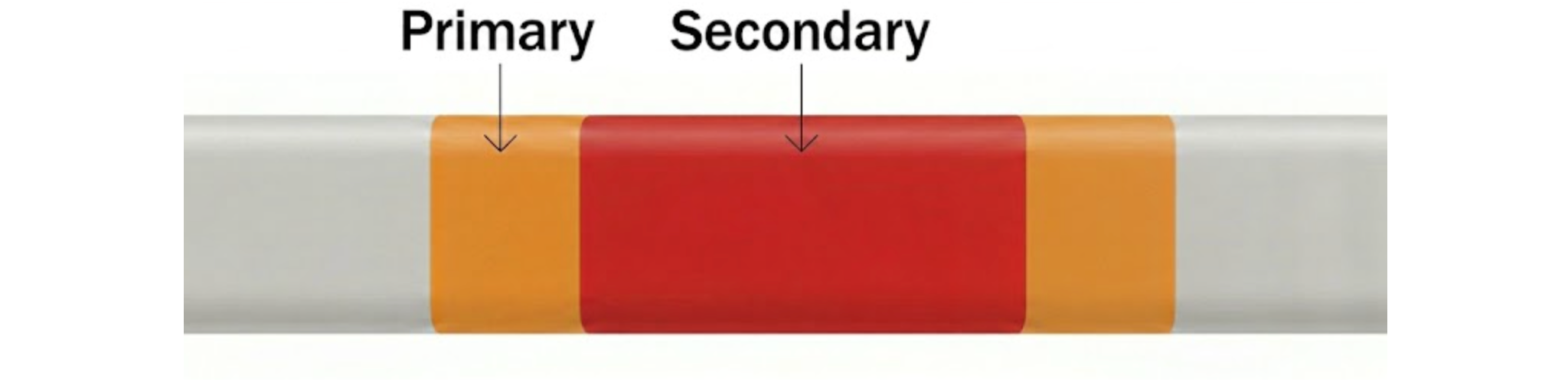}
    \caption{Front view of the wound bar.}
    \label{fig:barwounded}
\end{figure}

Consistent with typical experimental arrangements for testing magnetic materials, an excitation winding ($N_1$) of length $l_1$ is situated around the midpoint of the rod. A tightly coupled measurement coil ($N_2$), characterized by a length $l_2$, is then wound coaxially on top of the primary layer to ensure precise flux linkage detection. This tight configuration is practically achieved through the application of extremely fine winding wire. A visual representation of this entire coil geometry is provided in Fig.~\ref{fig:barwounded}.

Assuming the infinite solenoid approximation holds true, which requires $l_1 \gg a$ alongside measurement locations positioned far from the physical ends, the magnetic field impressed upon the surface of the core is assumed to be both spatially uniform and purely longitudinal.

\section{Modeling and Mathematical Proof}

\subsection{Assumptions and Operational Constraints}
The analytical framework for the square magnetic bar operates under the following key assumptions, which are direct extensions of the standard 1D cylindrical bar model \cite{BarC}:

\begin{itemize}
    \item \textbf{Infinite Solenoid Approximation:} To neglect longitudinal edge effects and fringing fields at the ends of the bar, it is assumed that the primary excitation winding is sufficiently long relative to the cross-section ($l_1 \gg a$). Consequently, the applied magnetic field at the surface of the core is considered uniform, purely axial (in the $z$-direction), and independent of the $z$-coordinate within the central measurement region.
    \item \textbf{Linearized Material Properties:} The magnetic hysteresis phenomenon is approximated using a localized linearization approach. At a given steady-state excitation amplitude and frequency, the core is assumed to exhibit a uniform, frequency-dependent complex permeability $\mu_c = \mu' - j\mu''$. This is valid when the magnetic field is sufficiently far from complete saturation.
    \item \textbf{Homogeneous Conductivity:} The electrical conductivity $\sigma$ of the magnetic steel is assumed to be homogeneous, constant, and isotropic throughout the core's volume.
    \item \textbf{Quasi-Static Regime:} The operational frequencies and cross-sectional dimensions are such that the electromagnetic wavelength is vastly larger than the bar dimensions. Therefore, the displacement current density ($\partial \mathbf{D}/\partial t$) is neglected in Ampère's law.
    \item \textbf{No-Load Secondary Circuit:} The voltage induced in the secondary coil ($V_2$) is measured using a high-impedance device (e.g., an oscilloscope). It is therefore assumed that the secondary current is zero ($I_2 = 0$), meaning the measurement circuit exerts no opposing magnetomotive force on the core.
\end{itemize}

\subsection{Fundamental Maxwell's Equations}
We consider time-harmonic fields ($e^{j\omega t}$) and operate under the quasi-static approximation where displacement currents are negligible. The governing Maxwell's equations inside the conductive magnetic bar are:
\begin{align}
    \nabla \times \mathbf{H} &= \sigma \mathbf{E} \label{eq:ampere} \\
    \nabla \times \mathbf{E} &= -j\omega \mathbf{B} \label{eq:faraday}
\end{align}

And we use the following material equation based on complex permeability:
\begin{equation}
    \mathbf{B} = \mu_c \mathbf{H}
\end{equation}
    
Driven by the solenoidal excitation, the straight bar's geometry ensures that the magnetic field is entirely axial, yielding $\mathbf{H} = H_z(x,y) \hat{\mathbf{z}}$. The induced electric fields and eddy currents circulate purely in the transverse $x-y$ plane.

We obtain the diffusion equation by taking the curl of Ampère's law and applying Faraday's law into the result:
\begin{align}
    \nabla \times (\nabla \times \mathbf{H}) &= -j\omega \mu_c \sigma \mathbf{H} \\
    \frac{\partial^2 H_z}{\partial x^2} + \frac{\partial^2 H_z}{\partial y^2} &= \gamma^2 H_z
\end{align}
where the propagation constant squared is $\gamma^2 = j\omega \mu_c \sigma$.

\subsection{Expressions for B and H via Separation of Variables}
By applying the method of separation of variables, the solution to this partial differential equation is:
\begin{equation}
    H_z(x,y) = X(x)Y(y)
\end{equation}
Using this formulation in the diffusion equation and dividing by $X(x)Y(y)$ gives:
\begin{equation}
    \frac{X''(x)}{X(x)} + \frac{Y''(y)}{Y(y)} = \gamma^2
\end{equation}
Due to the symmetry of the square cross-section, we have identical behavior along both the $x$ and $y$ axes. Therefore, considering each term equal to a constant $k^2$ yields:
\begin{equation}
    2k^2 = \gamma^2 \implies k = \sqrt{\frac{j\omega \mu_c \sigma}{2}}
\end{equation}

Given the symmetry of the core, the fields are required to be symmetric about the $x$- and $y$-axes. Consequently, the odd sinusoidal terms vanish, and the general solution reduces to:
\begin{equation}
    H_z(x,y) = H_0 \cosh(kx)\cosh(ky)
\end{equation}

The boundary condition at the surface of the bar is established by the primary excitation current $I$. Assuming a long solenoid, the surface field is:
\begin{equation}
    H_{surf} = \frac{N_1 I}{l_1}
\end{equation}
To determine the amplitude constant $H_0$, we match the field at the center of the boundary faces, e.g., at $(a/2, 0)$:
\begin{align}
    H_0 \cosh(k a/2) \cosh(0) &= H_{surf} \\
    H_0 &= \frac{H_{surf}}{\cosh(k a/2)}
\end{align}

This yields the exact analytical expressions for the internal fields:
\begin{equation}
    \boxed{H_z(x,y) = \frac{N_1 I}{l_1} \frac{\cosh(kx)\cosh(ky)}{\cosh(k a/2)}}
\end{equation}
\begin{equation}
    \boxed{B_z(x,y) = \mu_c \frac{N_1 I}{l_1} \frac{\cosh(kx)\cosh(ky)}{\cosh(k a/2)}}
\end{equation}

\subsection{Magnetic Flux Expression}
The total magnetic flux $\Phi$ through the cross-section is evaluated as the surface integral of $B_z$ as follows:
\begin{align}
    \Phi &= \int_{-a/2}^{a/2} \int_{-a/2}^{a/2} B_z(x,y) \,dx \,dy \\
    \Phi &= \mu_c \frac{H_{surf}}{\cosh(k a/2)} \left( \int_{-a/2}^{a/2} \cosh(kx) \,dx \right) \left( \int_{-a/2}^{a/2} \cosh(ky) \,dy \right) \\
    \Phi &= \mu_c \frac{H_{surf}}{\cosh(k a/2)} \left( \frac{2}{k} \sinh(k a/2) \right)^2
\end{align}
Substituting the expression for $H_{surf}$, the final expression for the total flux is:
\begin{equation}
    \boxed{\Phi = \mu_c \frac{N_1 I}{l_1} \frac{4}{k^2} \frac{\sinh^2(k a/2)}{\cosh(k a/2)}}
\end{equation}

\subsection{Mutual Complex Impedance}
By Faraday's law of induction, the induced electromotive force (voltage) in the open-circuited secondary winding is $V_2 = j\omega N_2 \Phi$. The mutual complex impedance $Z_m = \frac{V_2}{I}$ is:
\begin{align}
    Z_m &= j\omega N_2 \frac{\Phi}{I} \\
    Z_m &= j\omega \mu_c \frac{N_1 N_2}{l_1} \frac{4}{k^2} \frac{\sinh^2(k a/2)}{\cosh(k a/2)}
\end{align}
Substituting with the expression of the constant $k^2 = \frac{j\omega \mu_c \sigma}{2}$, the mutual impedance is expressed as:
\begin{equation}
    \boxed{Z_m = \frac{8 N_1 N_2}{\sigma l_1} \frac{\sinh^2(k a/2)}{\cosh(k a/2)}}
\end{equation}
For core loss evaluation, the equivalent primary self-impedance of the core is formulated as:
\begin{equation}
    Z_{core} = Z_m \frac{N_1}{N_2}
\end{equation}

\begin{equation}
    \boxed{Z_{core} = \frac{8 N_1^2}{\sigma l_1} \frac{\sinh^2(k a/2)}{\cosh(k a/2)}}
\end{equation}

\subsection{Loss Separation Expression}
To explicitly separate the losses, we map the core's equivalent primary admittance $Y_{core} = \frac{1}{Z_{core}} = G_{core} - jB_{core}$. The real part $G_{core}$ defines the active power in the core, corresponding to losses. 

To express separately the hysteresis loss, we evaluate the limit of the impedance when the core conductivity tends to zero ($\sigma \to 0$), thereby eliminating macroscopic eddy currents:
\begin{equation}
    Z_h = \lim_{\sigma \to 0} Z_{core} = j\omega \frac{N_1^2 a^2}{l_1} \mu_c
\end{equation}
The equivalent conductance representing only the intrinsic hysteresis is $G_h = \text{Re}(1/Z_h)$. 
The remaining conductance represents macroscopic eddy current losses: $G_e = G_{core} - G_h$.

Thus, the core power losses can be separated as:
\begin{equation}
    \boxed{P_h = |E_{1,rms}|^2 G_h}
\end{equation}

\begin{equation}
    \boxed{P_e = |E_{1,rms}|^2 G_e}
\end{equation}
with $E_{1,rms} = |Z_{core}| I_{rms}$.

\subsection{Apparent Permeability}
We can evaluate the impact of eddy currents on the energy-storage capability of the bar using the apparent permeability $\mu_{app}$, derived from the reactive part of the impedance. For an ideal core (no eddy currents, $\mu = \mu_{app}$), the imaginary part of the primary impedance is:
\begin{equation}
    \text{Im}(Z_{ideal}) = \omega \frac{N_1^2 a^2}{l_1} \mu_{app}
\end{equation}
Equating this to the imaginary part of the exact analytical complex impedance derived above:
\begin{equation}
    \omega \frac{N_1^2 a^2}{l_1} \mu_{app} = \text{Im} \left( \frac{8 N_1^2}{\sigma l_1} \frac{\sinh^2(k a/2)}{\cosh(k a/2)} \right)
\end{equation}
The apparent permeability $\mu_{app}$ is then expressed as:
\begin{equation}
    \boxed{\mu_{app} = \text{Im} \left( \frac{8}{\omega \sigma a^2} \frac{\sinh^2(k a/2)}{\cosh(k a/2)} \right)}
\end{equation}

\section{Summary of Main Formulas}
The fundamental analytical expressions defining the behavior of the square cross-section magnetic bar are summarized in Table~\ref{tab:summary}.

\renewcommand{\arraystretch}{1.9}
\begin{table}[h]
\centering
\caption{Summary of key analytical expressions.}
\label{tab:summary}
\begin{tabular}{p{5.5cm} p{10cm}}
\toprule
\textbf{Quantity} & \textbf{Expression} \\
\midrule
Magnetic field $H_z(x,y)$ & $\displaystyle H_z(x,y) = \frac{N_1 I}{l_1} \frac{\cosh(kx)\cosh(ky)}{\cosh(k a/2)}$ \\[1em]
Flux density $B_z(x,y)$ & $\displaystyle B_z(x,y) = \mu_c \frac{N_1 I}{l_1} \frac{\cosh(kx)\cosh(ky)}{\cosh(k a/2)}$ \\[1em]
Magnetic flux $\Phi$ & $\displaystyle \Phi = \mu_c \frac{N_1 I}{l_1} \frac{4}{k^2} \frac{\sinh^2(k a/2)}{\cosh(k a/2)}$ \\[1em]
Mutual impedance $Z_m$ & $\displaystyle Z_m = \frac{8 N_1 N_2}{\sigma l_1} \frac{\sinh^2(k a/2)}{\cosh(k a/2)}$ \\[1em]
Core impedance $Z_{core}$ & $\displaystyle Z_{core} = \frac{8 N_1^2}{\sigma l_1} \frac{\sinh^2(k a/2)}{\cosh(k a/2)}$ \\[1em]
Apparent permeability $\mu_{app}$ & $\displaystyle \mu_{app} = \text{Im} \left( \frac{8}{\omega \sigma a^2} \frac{\sinh^2(k a/2)}{\cosh(k a/2)} \right)$ \\[1em]
\bottomrule
\end{tabular}
\end{table}

\section{Results and Analysis}

This section presents the numerical results obtained from the analytical model derived for the square cross-section magnetic bar. The evaluation was conducted across a frequency sweep from 1~Hz to 1~MHz to characterize the material's behavior from the quasi-static regime up to frequencies where strong skin effects occur.

\subsection{Simulation Parameters}

The geometric dimensions, winding configuration, and material properties used in the simulation are summarized in Table~\ref{tab:sim_params}. It is noted that the electrical conductivity $\sigma$ is set to a relatively low value (90 S/m), which means significant skin effect phenomena are expected to manifest only at higher frequencies compared to typical highly conductive magnetic steels.

\renewcommand{\arraystretch}{1.4}
\begin{table}[h]
\centering
\caption{Simulation parameters for the square magnetic bar.}
\label{tab:sim_params}
\begin{tabular}{lccc}
\toprule
\textbf{Parameter} & \textbf{Symbol} & \textbf{Value} & \textbf{Unit} \\
\midrule
\textbf{Geometry} & & & \\
Square side length & $a$ & 10 & mm \\
Primary winding length & $l_1$ & 150 & mm \\
\textbf{Windings} & & & \\
Primary turns & $N_1$ & 1800 & - \\
Secondary turns & $N_2$ & 1350 & - \\
\textbf{Material \& Excitation} & & & \\
Intrinsic real relative permeability & $\mu'_r$ & 500 & - \\
Intrinsic imaginary relative permeability & $\mu''_r$ & 50 & - \\
Electrical conductivity & $\sigma$ & 90 & S/m \\
RMS Excitation current & $I_{rms}$ & 0.5 & A \\
Frequency range & $f$ & $10^0 - 10^6$ & Hz \\
\bottomrule
\end{tabular}
\end{table}

\subsection{Apparent Permeability Analysis}

\begin{figure}[h]
\centering
\includegraphics[width=0.99\linewidth]{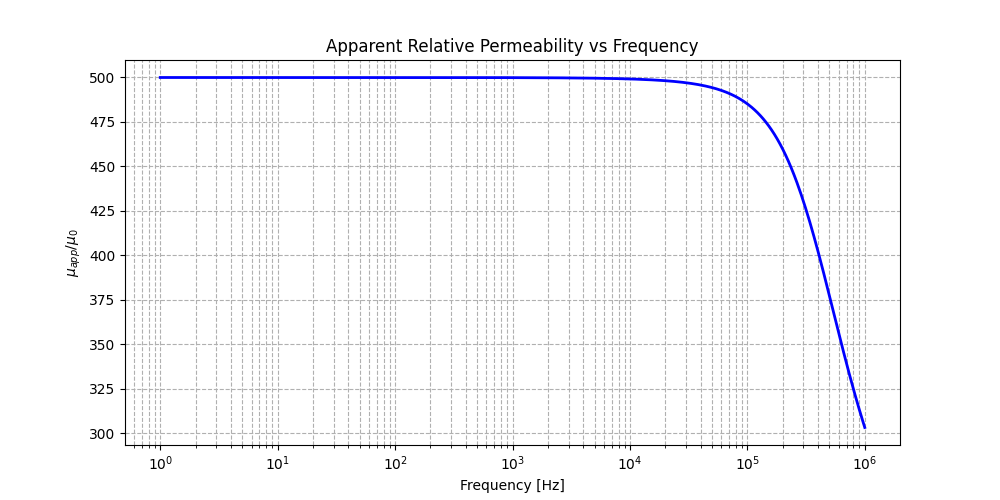}
\caption{Apparent relative permeability versus frequency.}
\label{fig:mu_app}
\end{figure}

Figure~\ref{fig:mu_app} illustrates the behavior of the apparent relative permeability ($\mu_{app}/\mu_0$) as a function of frequency.

At low frequencies (up to approximately 10~kHz), the apparent permeability remains constant at a value of 500. In this regime, the electromagnetic skin depth is significantly larger than the cross-sectional dimension $a$. Consequently, the magnetic flux penetrates the entire bar uniformly, and the apparent permeability equals the intrinsic real permeability of the material ($\mu_{app} \approx \mu'_r$).

As the frequency increases beyond 10~kHz, the skin effect becomes progressively significant. Induced eddy currents create opposing magnetic fields that shield the interior of the bar, forcing the magnetic flux to concentrate near the surface edges. This reduction in the effective magnetic cross-section causes the apparent permeability to decrease. At the maximum simulation frequency of 1~MHz, $\mu_{app}/\mu_0$ has decreased to approximately 300, indicating substantial magnetic shielding.

\subsection{Equivalent Core Impedance Analysis}

\begin{figure}[h]
\centering
\includegraphics[width=0.99\linewidth]{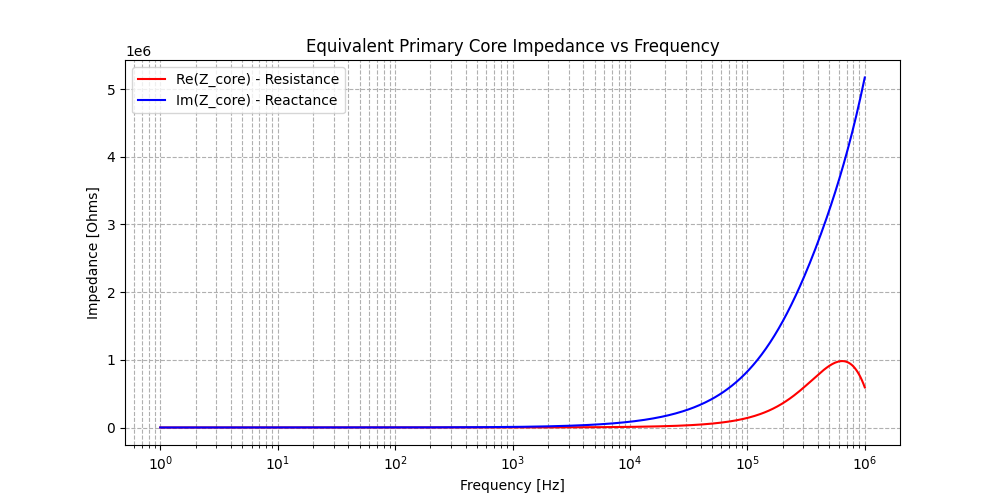}
\caption{Real and imaginary parts of the equivalent primary core impedance versus frequency.}
\label{fig:z_core}
\end{figure}

The equivalent primary impedance of the core, $Z_{core}$, is composed of a real part representing total core losses (equivalent resistance) and an imaginary part representing magnetic energy storage (inductive reactance). Their frequency dependence is shown in Figure~\ref{fig:z_core}.

The inductive reactance, $\text{Im}(Z_{core})$ (blue curve), dominates the impedance magnitude. It increases linearly with frequency in the low-frequency region, characteristic of an ideal inductor ($X_L = \omega L_{DC}$). At higher frequencies ($>100$ kHz), the curve deviates from linearity, bending upward due to the complex interplay between increasing frequency and decreasing apparent permeability caused by the skin effect.

The equivalent resistance, $\text{Re}(Z_{core})$ (red curve), represents the total power dissipated in the core ($P_{tot} = I_{rms}^2 \text{Re}(Z_{core})$). At low frequencies, the resistance is small and primarily due to hysteresis losses, which scale linearly with frequency. As frequency increases, eddy current losses, which scale with the square of the frequency, begin to dominate, causing the resistance to rise sharply after 10~kHz. The resistance reaches a peak near 700~kHz before slightly decreasing. This behavior is characteristic of analytical models employing hyperbolic functions; as the skin depth becomes much smaller than the conductor dimensions, the effective resistance transitions from a volumetric dependency ($\propto f^2$) to a surface-area dependency ($\propto \sqrt{f}$). The peak occurs at the transition point where the skin depth becomes comparable to the geometric dimensions, leading to a maximum in the real part of the impedance function.

\newpage
\bibliographystyle{ieeetr}
\bibliography{references}

\end{document}